\documentclass{article}
\usepackage{amssymb}
\usepackage{amsmath}
\usepackage[T1]{fontenc}
\newtheorem{theorem}{Theorem}
\newtheorem{acknowledgement}[theorem]{Acknowledgement}

\newtheorem{remark}[theorem]{Remark}

\begin{document}

\title{Asymptotically Flat Space -Times and its Hidden Recesses : \\
An Enigma from GR}
\author{ Ezra T. Newman}
\date{Feb.19, 2016}

\maketitle

\begin{abstract}
\ \ \ We begin by emphasizing that we are dealing with standard Einstein or
Einstein-Maxwell theory - absolutely no new physics has been inserted. The
fresh item is that the well-known asymptotically flat solutions of the
Einstein-Maxwell theory are transformed to a new coordinate system with
surprising and (seemingly) inexplicable results. We begin with the standard
description of (Null) Asymptotically Flat Space-Times described in
conventional Bondi-coordinates. After transforming the variables (mainly the
asymptotic Weyl tensor components) to a \textit{very special} set of NU
(Newman-Unti) coordinates, we find a series of relations totally mimicking
standard Newtonian classical mechanics and Maxwell theory. \ The surprising
and troubling aspect of these relations is that the associated motion and
radiation does \textit{not take place in physical space-time}. \ Instead
these relations takes place in an \textit{unusual inherited} complex
four-dimensional manifold referred to as H-Space that has no immediate
relationship with space-time. \ In fact these relations appear in two such
spaces, \textit{H}-Space and its dual space $\overline{H}$.\ 
\end{abstract}

\section{Introduction}

We begin by emphasizing that the material described here is based only on
standard Maxwell theory and classical general relativity. The matter and
charged sources are totally conventional - and absolutely no new physical
ideas have been introduced. \ \ The enigma lies in the fact that when a 
\emph{particular very special coordinate transformation }is applied to any
asymptotically flat (in the null sense, \cite{Bondi}\cite{Sachs}\cite{PandR}%
) solution of the Einstein-Maxwell equations in the neighborhood of future
null infinity (Scri,)\cite{NU}\cite{NT}, that had been given initially in
Bondi-Coordinates, certain very strange, seemingly inexplicable, relations
arise. First we obtain standard classical mechanic and classical Maxwell
theory relationships having nothing to do, in any obvious sense, with GR.
Among many more relations, described in detail later, we have, for example, $%
\overrightarrow{P}$=$M\overrightarrow{v}\ $or $\overrightarrow{L}$=$%
\overrightarrow{r}$x$\overrightarrow{P}\ \ $or the rate of quadrupole
radiation of angular momentum via a Maxwell field. There is even a special
case of $\overrightarrow{F}=M\overrightarrow{a}$. More startling is the fact
that these relations \underline{\textit{do not}}\textit{\ relate to
activities in physical space-time -} but rather to activities in an \textit{%
unusual inherited four-complex dimensional space, (a parameter space)}
referred to as $H$-space - and in fact, the activities occur in both $H$%
-space and/or in its dual $\overline{H}$-space\cite{KandN}. \ The $H$-space (%
$\overline{H}$-space) arises naturally via the solutions of a differential
equation associated with the coordinate transformation. We find these
results to be, at a minimum, surprising - and (so-far) inexplicable and/or
(maybe?)\ meaningless. They are also disturbing - why should GR produce such
strange relationships - without an underlying reason.

In Sec. II, we briefly review some of the well known results of
asymptotically flat space-times \cite{NP}\cite{NU}\cite{NT}\cite{LR}\cite%
{AandN1} while in Sec. III the special transformation from Bondi to the NU
coordinates\cite{NU} will be given\cite{LR}- essentially as a review. \ The
transformation is applied mainly to the Weyl tensor components (in
spin-coefficient form). The results of the transformation, i.e., the new
relations, are described in Sec. IV and followed by a discussion of these
relations in Sec.V and VI. \ 

The details of the transformation from Bondi to the special NU coordinates%
\cite{NU} are both long and complicated and have already been reported in
detail in a recent Living Review article. Rather than duplicate these
details we will, when needed, simply quote from the Living Review.\cite{LR}

The basic argument, though appearing to be rather complicated, has, for its
genesis, a simple analogue in the case of electro \& magneto-statics; the
determination of center of charge motion and the complex center of charge
motion. \ 

Consider a charge distribution, an origin and the associated electric dipole 
$\overrightarrow{D}_{E}$. \ A shift of origin by $\overrightarrow{R}$, so
that $\overrightarrow{r}$* = $\overrightarrow{r}-\overrightarrow{R}\ ,\ $%
leads to the dipole transformation, $\overrightarrow{D}_{E}^{\ast }=%
\overrightarrow{D}_{E}-q\overrightarrow{R}\ .\ $Setting $\overrightarrow{D}%
_{E}^{\ast }=0,\ $defines the center of charge by \ $\overrightarrow{R}=%
\overrightarrow{D}_{E}/q.\ $Formally this can be generalized to the complex
center of charge, by including the magnetic dipole,$\overrightarrow{D}_{M},$
via the complex dipole moment,$\ \overrightarrow{D}_{C}=\overrightarrow{D}%
_{E}+i\overrightarrow{D}_{M}.\ $Assume that it transforms under the complex
translation $\overrightarrow{r}$* = $\overrightarrow{r}-\overrightarrow{R}%
_{C},\ $as $\overrightarrow{D}_{C}^{\ast }=\overrightarrow{D}_{C}-q%
\overrightarrow{R}_{C}.\ $Setting $\overrightarrow{D}_{C}^{\ast }=0,$ we
have, by definition, 
\begin{equation}
\overrightarrow{R}_{C}=\frac{\overrightarrow{D}_{C}}{q},
\label{complex center of charge}
\end{equation}%
the position of the \textit{complex center of charge}.

This idea will be generalized to the asymptotically flat Einstein-Maxwell
fields where we will have, in addition to the \underline{complex}
electro-magnetic dipole, a generalization, by definition, to a \underline{%
complex} mass dipole (essentially, mass dipole $+\ i$ angular momentum).

From asymptotic information we will search for both the complex center of
mass and complex center of charge

The major issue will be what replaces the role of $\overrightarrow{R}_{C}.\ $%
The enigma then centers on the meaning of that replacement.

\section{Asymptotically Flat Space-Times}

As there is a great deal of existing literature on the asymptotic behavior
of the Einstein-Maxwell equations we will simply take what is needed from
this literature. \ Our two main sources are Newman-Penrose, (in
Scholarpedia) and Adamo-Newman (in Living Reviews). In all the discussions
we make heavy use of the NP formalism.

The study of asymptotically flat space-times was born from the early
brilliant work of H. Bondi\cite{Bondi} \cite{Sachs} where a one-parameter
family of null (i.e., characteristic) surfaces, $\mathfrak{C}_{u}$ labeled
by $u,$ was introduced as a (space-time) coordinate. Each of these surfaces
is generated by a two-parameter family of null geodesics, $\mathfrak{G,}$
each labeled by sphere coordinates $(\theta ,\phi )$ or equivalently (used
by us) by complex stereographic coordinates ($\zeta ,\overline{\zeta }$),
where $\zeta =e^{i\phi }\cot (\frac{\theta }{2}).$ The `length' along the
geodesics is given by the affine parameter, $r$.

The limit, $r=>\infty ,\ $is given as the null boundary of the compactified
space-time and referred to as Scri. It is coordinatized by ($u$,$\zeta ,%
\overline{\zeta }$). The full set of coordinates, ($u$,$\zeta ,\overline{%
\zeta },r$), called Bondi coordinates (Bondi himself did not use an affine
parameter, but what is referred to as a \textquotedblleft luminosity
distance\textquotedblright\ parameter), is not unique; there is a large
class of such coordinates. (We point out shortly that there is a
generalization of the Bondi coordinates, often referred to as NU coordinates%
\cite{NU}, a \ subset of which will shortly play a major role.) With the
choice of (any) one Bondi set, there is a natural choice of (Bondi) null
tetrad system, ( $l^{a},n^{a},m^{a},\overline{m}^{a}$). The vector $l^{a}$
is taken as the tangent vector to the null geodesics of $u$, $n^{a}\ $is
tangent to the null generators of Scri, while the ($m^{a},\overline{m}^{a}$)
are space-like tangent to Scri. \ They are normalized by all products
vanishing except $l^{a}n_{a}-1=m^{a}\overline{m}_{a}+1=0.\ $Their remaining
freedom is greatly limited by having the tetrad parallel propagated along
the null geodesics of\ $u$. When these restrictions are translated to the
tetrad, metric and the derivative operators become: 
\begin{eqnarray}
D &=&l^{a}\frac{\partial }{\partial x^{a}}=\frac{\partial }{\partial r}
\label{tet1} \\
\nabla &=&n^{a}\frac{\partial }{\partial x^{a}}=\frac{\partial }{\partial u}%
+U\frac{\partial }{\partial r}+X^{A}\frac{\partial }{\partial x^{A}}%
,~~~~~x^{A}=(x^{3},x^{4})=(\zeta ,\overline{\zeta }),  \label{tet2} \\
\delta &=&m^{a}\frac{\partial }{\partial x^{a}}=\omega \frac{\partial }{%
\partial r}+\xi ^{A}\frac{\partial }{\partial x^{A}},  \label{tet3} \\
\overline{\delta } &=&\overline{m}^{a}\frac{\partial }{\partial x^{a}}=%
\overline{\omega }\frac{\partial }{\partial r}+\overline{\xi }^{A}\frac{%
\partial }{\partial x^{A}}.  \label{tet4}
\end{eqnarray}%
\qquad\ 

The metric takes the form$\qquad \qquad $ 
\begin{equation}
g^{ab}=\left[ 
\begin{array}{ccc}
0\ , & 1\ , & 0 \\ 
1\ , & g^{22}\ , & g^{2A} \\ 
0\ , & g^{2A}\ , & g^{AB}%
\end{array}%
\right]  \label{g}
\end{equation}%
with 
\begin{eqnarray}
g^{22} &=&2(U-\omega \overline{\omega }),  \label{gab} \\
g^{2A} &=&X^{A}-(\overline{\omega }\xi ^{A}+\omega \overline{\xi }^{A}), 
\notag \\
g^{AB} &=&-(\xi ^{A}\overline{\xi }^{B}+\overline{\xi }^{A}\xi ^{B}),  \notag
\end{eqnarray}

Our major interest and concern will center on the Weyl and Maxwell tensors;
their asymptotic behavior, physical meaning, evolution and transformation
properties. \ We use the five complex self-dual NP components of the Weyl
tensor and three complex Maxwell components:\cite{NP}

\begin{eqnarray}
\Psi _{0} &=&-C_{abcd}l^{a}m^{b}l^{c}m^{d}=-C_{1313},  \label{W0} \\
\Psi _{1} &=&-C_{abcd}l^{a}n^{b}l^{c}m^{d}=-C_{1213},  \label{W1} \\
\Psi _{2} &=&-C_{abcd}l^{a}m^{b}\overline{m}^{c}n^{d}=-C_{1342},  \label{W2}
\\
\Psi _{3} &=&-C_{abcd}l^{a}n^{b}\overline{m}^{c}n^{d}=-C_{1242},  \label{W3}
\\
\Psi _{4} &=&-C_{abcd}n^{a}\overline{m}^{b}\overline{m}^{c}n^{d}=-C_{2442}.
\label{W4}
\end{eqnarray}

\begin{eqnarray*}
\phi _{0} &=&F_{ab}l^{a}m^{b}, \\
\phi _{1} &=&\frac{1}{2}F_{ab}(l^{a}n^{b}+m^{a}\overline{m}^{b}), \\
\phi _{2} &=&F_{ab}n^{a}\overline{m}^{b}.
\end{eqnarray*}

By integrating the radial asymptotic Bianchi identities and Maxwell
equations, we have what is known as the 'peeling' theorem\cite{NP}:%
\begin{eqnarray*}
\Psi _{0} &=&\Psi _{0}^{0}r^{-5}+O(r^{-6}), \\
\Psi _{1} &=&\Psi _{1}^{0}r^{-4}+O(r^{-5}), \\
\Psi _{2} &=&\Psi _{2}^{0}r^{-3}+O(r^{-4}), \\
\Psi _{3} &=&\Psi _{3}^{0}r^{-2}+O(r^{-3}), \\
\Psi _{4} &=&\Psi _{4}^{0}r^{-1}+O(r^{-2}).
\end{eqnarray*}%
\begin{eqnarray*}
\phi _{0} &=&\phi _{0}^{0}r^{-3}+O(r^{-4}), \\
\phi _{1} &=&\phi _{1}^{0}r^{-2}+O(r^{-3}), \\
\phi _{2} &=&\phi _{2}^{0}r^{-1}+O(r^{-2}),
\end{eqnarray*}%
with 
\begin{eqnarray*}
\Psi _{n}^{0} &=&\Psi _{n}^{0}(u,\zeta ,\overline{\zeta }), \\
\phi _{n}^{0} &=&\phi _{n}^{0}(u,\zeta ,\overline{\zeta }).
\end{eqnarray*}%
\qquad

The remaining (non-radial) Bianchi Identities and Maxwell equations yield
the evolution equations: \ 
\begin{eqnarray}
\dot{\Psi}_{2}^{0\,} &=&-\text{\dh }\Psi _{3}^{0\,}+\sigma ^{0}\Psi
_{4}^{0\,}+k\phi _{2}^{0}\overline{\phi }_{2}^{0},  \label{AsyBI1} \\
\dot{\Psi}_{1}^{0\,} &=&-\text{\dh }\Psi _{2}^{0\,}+2\sigma ^{0}\Psi
_{3}^{0\,}+2k\phi _{1}^{0}\overline{\phi }_{2}^{0},  \label{AsyBI2} \\
\dot{\Psi}_{0}^{0\,} &=&-\text{\dh }\Psi _{1}^{0\,}+3\sigma ^{0}\Psi
_{2}^{0\,}+3k\phi _{0}^{0}\overline{\phi }_{2}^{0},  \label{AsyBI3} \\
k &=&2Gc^{-4},
\end{eqnarray}

\begin{eqnarray}
\dot{\phi}_{1}^{0\,} &=&-\text{\dh }\phi _{2}^{0},  \label{MaxI} \\
\dot{\phi}_{0}^{0\,} &=&-\text{\dh }\phi _{1}^{0}+\sigma ^{0}\phi _{2}^{0}.
\label{MaxII}
\end{eqnarray}

These 5 equations, (\ref{AsyBI1}-\ref{MaxII}), after the coordinate
transformation to the special NU coordinates, contain our mechanical
equations of motion.

The quantity $\sigma ^{0}(u,\zeta ,\overline{\zeta })\ $is often called the
asymptotic shear, being the leading term in the shear of the geodesic
congruence, $l^{a};$\ i.e.$.,$%
\begin{equation*}
\sigma =r^{-2}\sigma ^{0}(u,\zeta ,\overline{\zeta })+O(r^{-4})
\end{equation*}%
while the first $u$-derivative of $\sigma ^{0}\ $is referred to as the Bondi
news function. \ We consider $\sigma ^{0}(u,\zeta ,\overline{\zeta })\ $as a
free function. It, as such, plays a significant role in what later follows.
From the spin-coefficient equations one finds that

\begin{eqnarray*}
\Psi _{3}^{0} &=&\text{\dh }(\overline{\sigma }^{0})^{\cdot }, \\
\Psi _{4}^{0} &=&-(\overline{\sigma }^{0})^{\cdot \cdot }\ 
\end{eqnarray*}

Defining the \textit{\ mass aspect,} ${\large \Psi ,}$ (real from field
equations) by%
\begin{equation}
\Psi =\overline{\Psi }\equiv \Psi _{2}^{0\,}+\eth ^{2}\overline{\sigma }%
^{0}+\sigma ^{0}(\overline{\sigma }^{0})^{\cdot },  \label{MassAspect}
\end{equation}

Bondi defines the asymptotic mass, $M_{B},$ and 3-momentum, $P_{B}^{i}\ \ $%
as the$\ l=0\ $\&$\ l=1\ $harmonic coefficients of $\Psi .\ $Specifically,

\textbf{Definition  1 \ }Identification of Physical Quantities:%
\begin{eqnarray}
\Psi  &=&\Psi ^{0}+\Psi ^{i}Y_{1i}^{0}+\Psi ^{ij}Y_{2ij}^{0}+.  \label{DEF.1}
\\
\Psi ^{0} &=&-\frac{2\sqrt{2}G}{c^{2}}M  \label{mass} \\
\Psi ^{i} &=&-\frac{6G}{c^{3}}P^{i}  \label{momentum}
\end{eqnarray}

By rewriting Eq.(\ref{AsyBI1}), replacing the $\Psi _{2}^{0\,}$ by $\Psi $
via Eq.(\ref{MassAspect}), we have

\begin{equation*}
\dot{\Psi}\text{ }=\text{ }(\sigma ^{0})^{\cdot }(\overline{\sigma }%
^{0})^{\cdot }+\ k\phi _{2}^{0}\overline{\phi }_{2}^{0},
\end{equation*}%
and one immediately has the Bondi mass/energy loss theorem:%
\begin{equation}
\dot{M}=-\frac{c^{2}}{2\sqrt{2}G}\int ((\sigma ^{0})^{\cdot }(\overline{%
\sigma }^{0})^{\cdot }+k\phi _{2}^{0}\overline{\phi }_{2}^{0})d^{2}S,
\label{BondiTheorem}
\end{equation}%
the integral taken over the unit 2-sphere. \ This relationship is at the
basis of all the contemporary work on the detection of gravitational
radiation.

\textbf{Definition\ \  2 \ \ }Though there has been no universal agreement,
we \underline{adopt the definition} of the \textit{complex} mass dipole
moment, $(D_{(complex)}^{i}=D_{(mass)}^{i}+ic^{-1}J^{i})$, as the $%
l=1\ $harmonic component of $\Psi _{1}^{0},\ $\ 
\begin{equation}
\Psi _{1}^{0}=-6\sqrt{2}Gc^{-2}(D_{(mass)}^{i}+ic^{-1}J^{i})Y_{1i}^{1}+....
\label{DEF.2}
\end{equation}%
$D^{i}\ $the mass dipole and $J^{i},\ $the total angular momentum, as seen
at null infinity.

\textbf{Definition\ \  3 \ }Our physical identification for the complex E\&M
dipole, (electric and magnetic dipoles, $(D_{Elec}^{i}+iD_{Mag})$) as the $%
l=1$ harmonic component of $\phi _{0}^{0}\ $is standard 
\begin{equation}
\phi _{0}^{0}=2(D_{Elec}^{i}+iD_{Mag})Y_{1i}^{1}.  \label{DEF.3}
\end{equation}%
\qquad 

Later we will connect these three physical identifications with the \textit{%
complex center of mass and the complex center of charge. }For the general
situation these two complex centers, are not yet defined and are independent.

\underline{ However, here, for simplicity, we will \textit{assume that they
coincide}}\emph{. \ This is not necessary but is a restriction.}

\section{Transformation To NU coordinates}

The Bondi coordinates (with their associated freedom, the BMS group\cite%
{Sachs}) are not the only useful coordinates to be used in the neigborhood
of Scri. \ Often the generalization, to what is often referred to as NU
coordinates, is called for. \ The basic idea of NU coordinates is to modify
the use of the uniform stacking of the constant $u$-slices of Scri and
instead allow for both the deformation of individual slices and the
arbitrary stacking of the new slices. \ Analytically the coordinate
transformation between Bondi, ($u,\zeta ,\overline{\zeta }$), and NU, ($\tau
,\zeta ,\overline{\zeta })$ coordinates is 
\begin{equation*}
u=G(\tau ,\zeta ,\overline{\zeta })
\end{equation*}%
with $G$ an arbitrary analytic function of $(\tau ,\zeta ,\overline{\zeta }%
).\ $Often we will allow the coordinates to be complex but close to the real
- which means that $\overline{\zeta }\ \ $can take values close to the $%
\emph{complex}\ $conjugate value\ of$\ \ \zeta .\ $This\ will\ be\ denoted\
by $\overline{\zeta }=>\widetilde{\zeta }.\ \ $The NU coordinates can be
thought of as being analogous to a coordinate system attached or associated
with an arbitrary world line. That analogy will shortly be taken a step
further.

The arbitrariness in the choice of $G\ \ $is greatly restricted by the
following argument. \ Null Geodesic Congruences (NGC) coming from the
interior of the space-time and intersecting with Scri are usually classified
by the values of the optical parameters near Scri, their divergence, shear
and twist. \ We now ask for \textit{slicings of Scri, \ (i.e., choices of }$%
G(\tau ,\zeta ,\overline{\zeta })$\textit{)} so that the NGCs normal to the
slicing have \emph{vanishing asymptotic shear}. \ The basic result from the
extensive literature (\cite{Good Cut Eq}, \cite{LR}\emph{) }is that $G\ $%
must satisfy the so-called good-cut equation, 
\begin{equation}
\text{\dh }^{2}G=\sigma ^{0}(G,\zeta ,\overline{\zeta }).
\label{good cut Eq}
\end{equation}

With regularity conditions, the solution space for this equation is a
four-complex dimensional parameter space, $z^{a},\ $referred to as H-Space%
\cite{NP2}\cite{H}. \ Solutions, with appropriate choice of H-space
coordinates, can be written as 
\begin{eqnarray}
u &=&Z(z^{a},\zeta ,\overline{\zeta })=z^{a}l_{a}^{\#}(\zeta ,\overline{%
\zeta })+\Sigma _{l\geq 2}Z_{lm}(z^{a})Y_{lm}(\zeta ,\overline{\zeta })
\label{Z} \\
l_{a}^{\#}(\zeta ,\overline{\zeta }) &=&(\frac{\sqrt{2}}{2}Y_{00},\frac{1}{2}%
Y_{1m})=\frac{\sqrt{2}}{2(1+\zeta \overline{\zeta })}(1+\zeta \overline{%
\zeta },\zeta +\overline{\zeta },i\overline{\zeta }-i\zeta ,-1+\zeta 
\overline{\zeta })  \label{l2}
\end{eqnarray}

\ H-Space plays several roles. First of all, it possesses a natural complex
metric that automatically is Ricci flat and anti-self dual\cite{NP2}.
Penrose uses the H-Space for his non-linear graviton construction. \ In the
metric context there are strong suggestions that $H$-space$\ $and its
properties\ are closely related to the ideas of the present work. The
details, however, are not yet clear. \ We use H-Space without reference to
its metric.

We choose an arbitrary (for the time being) one-complex dimensional curve, $%
\ z^{a}=\xi ^{a}(\tau ),\ $and insert it into Eq.(\ref{Z}) so that we get a
one parameter family of cuts%
\begin{equation}
u=Z(\xi ^{a}(\tau ),\zeta ,\overline{\zeta })=G(\tau ,\zeta ,\overline{\zeta 
})  \label{solution}
\end{equation}%
depending on choice of the world line. \ The solutions (with an appropriate
choice of H-Space coordinates) has the form%
\begin{equation*}
u=G(\tau ,\zeta ,\overline{\zeta })=\xi ^{a}(\tau )l_{a}^{\#}(\zeta ,%
\overline{\zeta })+\Sigma _{l\geq 2}Z_{lm}(\xi ^{a}(\tau ))Y_{lm}(\zeta ,%
\overline{\zeta }).
\end{equation*}

This \emph{world-line in H-space} will become our complex center of mass and
center of charge, the analogue of the electro-magnetic, Eq.(\ref{complex
center of charge}). The inverse function is written as

\begin{equation}
\tau =T(u,\zeta ,\overline{\zeta }).  \label{T}
\end{equation}

Note that though $G(\tau ,\zeta ,\overline{\zeta })\ $is in general complex
one can construct from $G,\ $a one parameter family of real cuts by taking
the real part of $G$.

\subsection{The Transformation{\protect\LARGE \ }}

Our task is to transform the Weyl tensor\textit{\ physical identification},
i.e., the complex mass (or gravitational) dipole, Eq.(\ref{DEF.2}), from the
Bondi coordinates ($u=$constant slicings) to the NU coordinates ($\tau $%
=constant slicings).\ This is \emph{a very tedious and long task} - done by
approximations and Clebsch-Gordon expansions - that is only outlined here%
\cite{LR}. When this complex mass (or gravitational) dipole is expressed on
the $\tau \ $slices, we set it (and, by assumption, the complex E\&M dipole)
to zero. This determines the world-line,$\ \ z^{a}=\xi ^{a}(\tau ),\ $which, 
\emph{by definition}, is our complex center of mass world-line (and, by our
assumption, our complex center of charge).

\begin{remark}
\textit{At this point there is no reason to take this definition of a
"complex world-line" seriously or for \vspace{0in}\nolinebreak it to have
any physical significance. The "world-line" seems to have no relationship
with space-time points. It is defined on the rather mysterious four-complex
dimensional }$H$\textit{-space. (The Penrose non-linear graviton.)
Nevertheless, in the following section we will see that in fact it does have
a very close relationship with things of obvious great physical significance.%
}
\end{remark}

\textit{It is the fact of the existence of these relationships that is our
enigma - why do they exist?}

The transformation of the complex mass dipole from $u$-coordinates to $\tau $%
-coordinates is, as we said, long and painful. It begins with the null
rotation of the Bondi tetrad [based on the (Bondi) null vector, $l^{a},$%
normal to the Bondi slices] to the tetrad based on the null vector $l^{\ast
a},\ ($normal to the $\tau $-slices);

\begin{eqnarray}
l^{\ast a} &=&l^{a}+b\overline{m}^{a}+\overline{b}m^{a}+b\overline{b}n^{a}
\label{null rotation} \\
m^{\ast a} &=&m^{a}+bn^{a}  \notag \\
n^{\ast a} &=&n^{a}  \notag
\end{eqnarray}%
where 
\begin{equation*}
b=-\ r^{-1}{\large L+O(r^{-2}),}
\end{equation*}%
and\ $L,\ $the angle field at Scri between $l^{\ast a}\ $and $l^{a},$

\begin{equation*}
{\large L(u,\zeta ,\overline{\zeta })=\eth G(\tau ,\zeta ,\overline{\zeta })|%
}_{{\large \tau =T(u,\zeta ,\overline{\zeta })}}{\large .}
\end{equation*}%
\qquad

From the null rotation we find that the asymptotic Weyl tensor components
transform as\cite{LR},

\begin{eqnarray}
\Psi _{0}^{\ast 0} &=&\Psi _{0}^{0}-4L\Psi _{1}^{0}+6L^{2}\Psi
_{2}^{0}-4L^{3}\Psi _{3}^{0}+L^{4}\Psi _{4}^{0},  \label{0} \\
\Psi _{1}^{\ast 0} &=&\Psi _{1}^{0}-3L\Psi _{2}^{0}+3L^{2}\Psi
_{3}^{0}-L^{3}\Psi _{4}^{0},  \label{1} \\
\Psi _{2}^{\ast 0} &=&\Psi _{2}^{0}-2L\Psi _{3}^{0}+L^{2}\Psi _{4}^{0},
\label{2} \\
\Psi _{3}^{\ast 0} &=&\Psi _{3}^{0}-L\Psi _{4}^{0},  \label{3} \\
\Psi _{4}^{\ast 0} &=&\Psi _{4}^{0}.  \label{4}
\end{eqnarray}

Our procedure for finding the complex center of mass now centers on Eq.(\ref%
{1}). We search for and \underline{\emph{set to zero}} the$\ l=1\ $spherical
harmonic coefficient of $\Psi _{1}^{\ast 0}\ $on a constant $\tau $ slice. \
The right-side of Eq.(\ref{1}), a function of $u,$ is first converted to a
function of $\tau \ $via Eq.(\ref{solution}).$\ \ $All the variables on the
right side are then expanded in spherical harmonics and simplified by
Clebsch-Gordon expansions. \ The isolation of the $l=1\ $harmonics is still
difficult and approximations - to second order in the variables - are needed.

The result from these operations on Eq.(\ref{1}) leads to an expression for
the $l=1$ coefficients of Bondi $\Psi _{1}^{0},\ $namely the following: 
\begin{eqnarray*}
\Psi _{1}^{0i} &=&-\frac{6\sqrt{2}G}{c^{2}}M_{B}\xi ^{i}+i\frac{6\sqrt{2}G}{%
c^{3}}P^{k}\xi ^{j}\epsilon _{kji}-\frac{576G}{5c^{3}}P^{k}\xi ^{ik}+i\frac{%
6912\sqrt{2}}{5}\xi ^{lj}\overline{\xi }^{lk}\epsilon _{jki} \\
&&-i\frac{2\sqrt{2}G}{c^{6}}q^{2}\xi ^{k}\overline{\xi }^{j\prime \prime
}\epsilon _{kji}-\frac{48G}{5c^{6}}q^{2}\xi ^{ji}\overline{\xi }^{j\prime
\prime }-\frac{4G}{5c^{7}}q^{2}\xi ^{j}\overline{Q}_{C}^{ij\prime \prime
\prime }-i\frac{16\sqrt{2}G}{5c^{7}}q\xi ^{lj}\overline{Q}_{C}^{lk\prime
\prime \prime }\epsilon _{jki}.
\end{eqnarray*}%
where%
\begin{equation*}
\xi ^{ij}=\frac{\sqrt{2}G}{24c^{4}}(Q_{Mass}^{ij\prime \prime
}+iQ_{spin}^{ij\prime \prime })
\end{equation*}%
are the time-derivatives of the gravitational quadrupoles and $Q_{C}^{ij}$\
the electric and magnetic quadrupoles.

Treating the quadrupoles and higher powers of $c^{-1}\ $as small, writing

\begin{equation}
\xi ^{i}=\xi _{R}^{i}+i\xi _{I}^{i},  \label{real&complex}
\end{equation}%
and remembering our physical identifications, Eq.(\ref{DEF.2}), we are left
with

\begin{eqnarray}
D_{(mass)}^{i} &=&M_{B}\xi _{R}^{i}-c^{-1}P^{k}\xi _{I}^{j}\ \epsilon
_{jki}+...,  \label{mass  dipole} \\
J^{i} &=&cM_{B}\xi _{I}^{i}+P^{k}\xi _{R}^{j}\epsilon _{jki}+....
\label{ang mom}
\end{eqnarray}%
\qquad

In other words we have the conventional definition of mass dipole (M$\ 
\overrightarrow{\mathbf{r}})\ $augmented with an unusual term, $%
\overrightarrow{P}\mathrm{x}\overrightarrow{S},$ discussed later, i.e.,%
\begin{equation}
\overrightarrow{D}_{(mass)}=M_{B}\overrightarrow{r}+c^{-2}M_{B}^{-1}%
\overrightarrow{P}\mathrm{x}\overrightarrow{S}.  \label{D}
\end{equation}
\ 

In addition we have an expression for angular momentum; the intrinsic spin $%
\overrightarrow{S}$ (same as for the Kerr metric) plus the orbital angular
momentum 
\begin{equation}
\overrightarrow{J}=\overrightarrow{S}+\overrightarrow{r}\mathrm{x}%
\overrightarrow{P}=cM_{B}\overrightarrow{\xi }_{I}+\overrightarrow{r}\mathrm{%
x}\overrightarrow{P}.  \label{J}
\end{equation}

We find this to be a rather startling result; standard classical mechanical
relationships where ordinary space-time has been replaced (not by any
assumption but automatically) by the mysterious H-space. \ It's our first
set of enigmas. \ We return to them later in connection with the
relativistic angular momentum tensor and the $\overrightarrow{P}\mathrm{x}%
\overrightarrow{S}\ $term.

Further classical mechanical relations, even more startling, are given in
the next section.

\section{More Results}

The relations for the mass dipole and the angular momentum, given by Eqs. (%
\ref{mass dipole}) \& (\ref{ang mom}) are now substituted into the
evolutionary Bianchi Identities Eqs.(\ref{AsyBI2}), (\ref{AsyBI3}) and
Maxwell Eq.(\ref{MaxI}) and (\ref{MaxII})%
\begin{eqnarray}
\dot{\Psi}_{2}^{0\,} &=&-\text{\dh }\Psi _{3}^{0\,}+\sigma ^{0}\Psi
_{4}^{0\,}+k\phi _{2}^{0}\overline{\phi }_{2}^{0},  \label{BI1} \\
\dot{\Psi}_{1}^{0\,} &=&-\text{\dh }\Psi _{2}^{0\,}+2\sigma ^{0}\Psi
_{3}^{0\,}+2k\phi _{1}^{0}\overline{\phi }_{2}^{0},  \label{BI2}
\end{eqnarray}

\begin{eqnarray}
\dot{\phi}_{1}^{0\,} &=&-\text{\dh }\phi _{2}^{0}  \label{M1} \\
\dot{\phi}_{0}^{0\,} &=&-\text{\dh }\phi _{1}^{0}+\sigma ^{0}\phi _{2}^{0}
\label{M2}
\end{eqnarray}

From the real part of Eq.(\ref{BI2}) we get an expression for the Bondi
momentum in terms of the time derivative of the mass dipole;%
\begin{eqnarray}
P^{i} &=&M_{B}\xi _{R}^{i\prime }-\frac{2q^{2}}{3c^{3}}\xi _{R}^{i\prime
\prime }+H.O.  \label{Mv} \\
H.O. &=&\text{higher order terms}  \notag
\end{eqnarray}%
i.e., we obtain, for the momentum, the kinematic $M\overrightarrow{v}$ term
and a term familiar from electrodynamics, the radiation reaction
contribution to the linear momentum.

From the imaginary part of the Bianchi Identity we have the momentum loss
equation;%
\begin{equation}
J^{i\prime }=-\frac{2q^{2}}{3c^{3}}\xi _{I}^{i\prime \prime }+\frac{2q^{2}}{%
3c^{3}}(\xi _{R}^{j\prime }\xi _{R}^{k\prime \prime }+\xi _{I}^{k\prime }\xi
_{I}^{k\prime \prime })\epsilon _{kji}+\text{Mass\&E\&M quadrupole terms.}
\label{J'}
\end{equation}

\emph{To our knowledge this might be a new expression for angular momentum
loss. We have not found anything like this in the literature.}

Going to the next Bianchi identity, Eq(\ref{BI1}), we get relations for both
the $l=0\ $and the$\ l=1$\ harmonic terms.

First we have the (Bondi) mass loss expression but now augmented by the
electromagnetic loses.%
\begin{eqnarray}
M_{B}^{\prime } &=&-\frac{G}{5c^{7}}(Q_{Mass}^{jk\prime \prime \prime
}Q_{Mass}^{jk\prime \prime \prime }+Q_{Spin}^{jk\prime \prime \prime
}Q_{Spin}^{jk\prime \prime \prime })-\frac{4q^{2}}{3c^{5}}(\xi _{R}^{i\prime
\prime }\xi _{R}^{i\prime \prime }+\xi _{I}^{i\prime \prime }\xi
_{I}^{i\prime \prime })  \label{mass loss} \\
&&-\frac{4}{45c^{7}}(Q_{E}^{jk\prime \prime \prime }Q_{E}^{jk\prime \prime
\prime }+Q_{M}^{jk\prime \prime \prime }Q_{M}^{jk\prime \prime \prime }), 
\notag
\end{eqnarray}%
where the first term is the standard Bondi quadrupole mass loss (including
now the spin quadrupole contribution to the loss), the second term and third
terms are the standard E\ \&\ M dipole and quadrupole energy loss.

From the $l=1\ $terms we get the momentum loss

\begin{equation}
P^{i\prime }=F_{recoil}^{i}  \label{P'}
\end{equation}%
where $F_{recoil}^{i}\ $is composed of many non-linear radiation terms
involving the time derivatives of the gravitational quadrupole and the E\&M
dipole and quadrupole moments whose details are not now relevant for us. \
We however can substitute Eq.(\ref{Mv}) into Eq.(\ref{P'}) which leads
Newtons second law;

\begin{equation}
M_{B}\xi _{R}^{i\prime \prime }=F^{i}=M_{B}^{\prime }\xi _{R}^{i\prime }+%
\frac{2q^{2}}{3c^{3}}\xi _{R}^{i\prime \prime \prime }+F_{recoil}^{i}.
\label{F=ma}
\end{equation}%
The first force term, on the right, is the standard rocket mass loss
expression while the second is the "famous" radiation reaction force of
classical Maxwell theory.

There are two further items worth mentioning.

1. From our earlier results,

i.$\ \ \ \ \ \xi _{R}^{i}=$center of mass position

ii. \ \ \ $S^{i}=Mc\xi _{I}^{i}\ $=\ Spin-Angular momentum

iii. \ \ \ $D_{M}^{i}=q\xi _{I}^{i}=\ $Magnetic dipole Moment

We have, from comparing the classical gyromagnetic ratio $\gamma =\frac{q}{%
2Mc}g\ ,\ $with ours 
\begin{equation}
\gamma =\frac{D_{M}^{i}}{L_{spinang.mom}}=\frac{q\xi _{I}^{i}}{M_{B}c\xi
_{I}^{i}}=\frac{q}{M_{B}c},  \label{g factor}
\end{equation}%
we find the Dirac value of the $g$-factor, i.e., $g$=$2$.

2. In classical relativistic mechanics one defines the angular momentum
tensor, $M^{ab}$,

\begin{eqnarray*}
M^{ab} &=&L^{ab}+S^{ab} \\
L^{ab} &=&2MX^{[a}V^{b]} \\
S^{ab} &=&-\eta ^{abcd}S_{c}^{\ast }V_{d},\ \ \ S_{c}^{\ast }V^{c}=0
\end{eqnarray*}%
so that%
\begin{eqnarray}
M^{ij} &=&L^{ij}+S^{ij}  \label{ang mom 2} \\
&=&M(X^{i}V^{j}-V^{i}X^{j})-\epsilon ^{ijk}(S_{k}^{\ast
}V_{0}-V_{k}S_{0}^{\ast })  \notag
\end{eqnarray}

Using our results, $S^{i}=cS^{\ast i}=Mc\xi _{I}^{i},\ S_{0}=0,\ V_{0}\sim
1,\ V_{k}\sim 0\ $and multiplying by $\epsilon _{ijk},\ $we have agreement
with our Eq.(\ref{ang mom}).

Then from%
\begin{eqnarray}
M^{0i} &=&L^{0i}+S^{0i}  \label{mass dipole 2} \\
&=&2MX^{[0}V^{i]}-\eta ^{0ijk}S_{j}^{\ast }V_{k}  \notag \\
&=&M_{B}\xi _{R}^{i}-\epsilon _{ijk}c^{-1}\xi _{I}^{j}P^{k},  \notag
\end{eqnarray}

\noindent we have agreement with our Eq.(\ref{mass dipole}), i.e., with our $%
\overrightarrow{P}\mathrm{x}\overrightarrow{S}\ $term.

The \textit{relativistic angular momentum tensor} (unrelated to physical
space-time) is sitting in our Weyl tensor.

It is the collection of results, Eqs. (\ref{mass dipole}), (\ref{ang mom}), (%
\ref{Mv}),(\ref{J'}),(\ref{mass loss}) and (\ref{F=ma}), mimicking or
imitating classical mechanics, that constitute our enigma. \ Though they
appear to be space-time equations of motion there is \emph{no space-time
that is associated with the equations}. They (as we said earlier) take place
on the strange H-space. \ Is this just a giant coincidence? That is
difficult to believe. \ What possible meaning can one give to them - and to
the H-space? \ What is the \ physical meaning of $\xi ^{i}(\tau )=\xi
_{R}^{i}(\tau )+i\xi _{I}^{i}(\tau ).$

As a final comment we point out that starting with the same asymptotically
flat space time but now looking for the conjugate slicing of its Scri, i.e.,
looking for the slicing that is associated with \textit{complex conjugate }%
shear-free null surfaces yields at alternate (conjugate) NU coordinate
system and a construction of the conjugate classical mechanical relations.
We obtain instead of $H$-space, the conjugate $\overline{H}$-space. In other
words there is a dual description of everything that was described here.

\section{An Example}

If we consider the very special (radiationless) case where the Bondi
asymptotic shear, vanishes, i.e., $\sigma ^{0}(u,\zeta .\overline{\zeta }%
)=0,\ \cite{AandN2}$the Good Cut Equation simplifies to%
\begin{equation}
\text{\dh }^{2}G=0,  \label{sigma=0}
\end{equation}%
with regular solutions being 
\begin{eqnarray}
u &=&Z(z^{a},\zeta ,\overline{\zeta })=z^{a}l_{a}^{\#}(\zeta ,\overline{%
\zeta }),  \label{Z1} \\
l_{a}^{\#}(\zeta ,\overline{\zeta }) &=&(\frac{\sqrt{2}}{2}Y_{00},\frac{1}{2}%
Y_{1m})=\frac{\sqrt{2}}{2(1+\zeta \overline{\zeta })}(1+\zeta \overline{%
\zeta },\zeta +\overline{\zeta },i\overline{\zeta }-i\zeta ,-1+\zeta 
\overline{\zeta }).  \label{l*}
\end{eqnarray}

The four complex dimensional solutions space, $z^{a},\ i.e,.$the H-space,
becomes in this case 'complex Minkowski space", with metric ds$^{2}=\eta
_{ab}dz^{a}dz^{b}\ \cite{LR}.\ $Choosing a one-complex dimensional curve, $\
z^{a}=\xi ^{a}(\tau ),\ $and inserting it into Eq.(\ref{Z1}), we get a one
parameter family of cuts%
\begin{equation}
u=\xi ^{a}(\tau )l_{a}^{\#}(\zeta ,\overline{\zeta }),
\end{equation}%
that mimic the cuts of Scri of flat space.

All our equations of motion can now be thought of as taking place in
Minkowski space.

Our kinematic equations, \ref{D} and \ref{J},

\begin{equation}
\overrightarrow{D}_{(mass)}=M_{B}\overrightarrow{r}+c^{-2}M_{B}^{-1}%
\overrightarrow{P}\mathrm{x}\overrightarrow{S}.
\end{equation}%
\ 

\begin{equation}
\overrightarrow{J}=\overrightarrow{S}+\overrightarrow{r}\mathrm{x}%
\overrightarrow{P}=cM_{B}\overrightarrow{\xi }_{I}+\overrightarrow{r}\mathrm{%
x}\overrightarrow{P}.
\end{equation}%
are now 3-D flat space relations.

\section{Discussion \& Comments}

We want in this section to reiterate and emphasize the strangeness of the
results presented.

a. In the search of null surfaces that resemble, at infinity, ordinary
Minkowski light-cones - i.e., shear-free null and diverging (as $r^{-1})$
surfaces - we find a four complex parameter family of such surfaces,
H-space. \ This family DOES\ NOT\ CONSTITUTE THE POINTS OF SPACE-TIME. \ \
Nevertheless we find that much of classical mechanics seems to live on
H-space. \ If the starting space-time had been flat, the H-space would have
been complex Minkowski space \ and its real part could be identified with
the physical space-time. \ One might think of H-space as an observation or
'Mirage' space - though this does not appear to clarify its physical meaning.

b. If we are dealing with asymptotic Einstein-Maxwell, the H-space equations
of motion contain the classical Radiation Reaction term without the standard
use of mass renormalization - Just \ the "Asymptotic" Bondi Mass.

c. If (and we do not know if it is true) the Einstein-Maxwell Solutions are
stable - then something in the equations suppresses radiation reaction run
away behavior? It is easily seen that the equations of motion of a charged
particle contain not only the radiation reaction force but other
non-standard terms\ref{F=ma}. What is the effect of these terms?

d. One can see the details in some special cases - e.g., all asymptotic
Static and Stationary metrics and the Robinson-Trautman class\cite{LR}.

e. For no obvious reason, one obtains the Dirac value of the gyromagnetic
ratio. This is related to the assumption made at the start that the complex
center of mass coincides with the complex electric dipole moment.

f. It definitely appears as if there are further structures to be explored.
They are related to the question of the real and imaginary parts of the
shear-free NU cuts of Scri.

\section{Questions}

a. We feel that these results might be profound but we are stuck how to
further interpret or further use them. \ On Tuesdays, Thursdays, Saturdays
and Sundays we are rather certain that they are profound but unfortunately
on Mondays, Wednesdays and Fridays we wonder if there is anything of
substance?

b. We have the "motion" of center of mass of the entire system. \ Can one do
two body Equations of Motion? How? It is not clear but there
seems to be a real possibility to do this by treating the space between two
distantly separated masses as approximately flat and behaving similarly to
Scri.

c. How do these results enter into the issues of Quantum Gravity - or even
ordinary Quantum Mechanics? \ Perhaps on the profound side?

d. Our complex world-line sits in H-Space whose metric satisfies the COMPLEX
Einstein Eqs. What does this mean? \ Is there any further significance of
this?

e. H-Space plays a fundamental role in Penrose's Non-Linear Graviton
construction? \cite{P} So What? Is this just a fact with no further
significance?

f. \ Is our present enigma at all related to the so-called holographic
principle\cite{AandN2}?

g. If these H-Space equations of motion were to contain potentially new
phenomena, do we take that seriously and look for experimental confirmation?

\begin{acknowledgement}
\end{acknowledgement}

We thank Timothy Adamo for both a careful critical reading of the
manuscript, for hours of enlightening discussions and collaboration on an
earlier manuscript where many of the present ideas were developed.

\end{document}